\headline={\ifnum\pageno=1\firstheadline\else
\ifodd\pageno\rightheadline \else\leftheadline\fi\fi}
\def\firstheadline{\hfil}
\def\rightheadline{\hfil}
\def\leftheadline{\hfil}

\font\twelvebf=cmbx10 scaled\magstep 1
\font\twelverm=cmr10 scaled\magstep 1
\font\twelveit=cmti10 scaled\magstep 1

\font\tenbf=cmbx10
\font\tenrm=cmr10
\font\tenit=cmti10

\font\ninerm=cmr9

\parindent=1.5pc
\hsize=6.0truein
\vsize=8.5truein
\nopagenumbers

\def\ag{{\cal A}/{\cal G}}
\def\agb{\overline{\ag}}
\def\a{\cal A}
\def\g{\cal G}
\def\Cyl{{\rm Cyl}}
\def\HA{\overline{\cal HA}}
\def\Ab{\bar{A}}
\def\tit{\twelveit}
\def\tbf{\twelvebf}

\def\ut{\underline}

\centerline{\tenbf POLYMER GEOMETRY AT PLANCK SCALE}
\centerline{\tenbf AND QUANTUM EINSTEIN EQUATIONS}
\baselineskip=16pt
\vglue 0.8cm
\centerline{\tenrm ABHAY ASHTEKAR}
\baselineskip=13pt
\centerline{\tenit Center for Gravitational Physics and Geometry}
\baselineskip=12pt
\centerline{\tenit Physics Department, Penn State, University Park, 
PA 16802}
\vglue 0.3cm
\vglue 0.8cm
\centerline{\tenrm ABSTRACT}
\vglue 0.3cm
{\rightskip=3pc \leftskip=3pc \tenrm\baselineskip=12pt\noindent Over
the last two years, the canonical approach to quantum gravity based on
connections and triads has been put on a firm mathematical footing
through the development and application of a new functional calculus
on the space of gauge equivalent connections. This calculus does not
use any background fields (such as a metric) and is thus well-suited to a
fully non-perturbative treatment of quantum gravity. Using this
framework, quantum geometry is examined. Fundamental excitations turn
out to be one-dimensional, rather like polymers. Geometrical
observables such as areas of surfaces and volumes of regions have
purely discrete spectra. Continuum picture arises only upon coarse
graining of suitable semi-classical states. Next, regulated quantum
diffeomorphism constraints can be imposed in an anomaly-free fashion
and the space of solutions can be given a natural Hilbert space
structure.  Progress has also been made on the quantum Hamiltonian
constraint in a number of directions. In particular, there is a recent
approach based on a generalized Wick transformation which maps
solutions to the Euclidean quantum constraints to those of the
Lorentzian theory. These developments are summarized. Emphasis is on
conveying the underlying ideas and overall pictures rather than
technical details.}
\vglue 0.6cm
\vfil
\twelverm
\baselineskip=14pt
\leftline{\twelvebf 1. Introduction}
\vglue 0.4cm

It is well known that quantum general relativity is perturbatively
non-re\-norma\-lizable. Particle theorists often take this to be a
sufficient reason to abandon general relativity and seek an
alternative which has a better ultraviolet behavior in perturbation
theory. However, one is by no means forced to this route. For, there
do exist a number of field theories which are perturbatively
non-renormalizable but are {\it exactly soluble}. An outstanding
example is the Gross-Neveau model in 3 dimensions, $(GN)_3$, which was
recently shown to be exactly soluble rigorously [1]. Furthermore, the
model does not exhibit any mathematical pathologies. For example, it
was at first conjectured that the Wightman functions of a
non-renormalizable theory would have a worse mathematical behavior.
The solution to $(GN)_3$ showed that this is not the case; as in
familiar renormalizable theories, they are tempered distributions.
Thus, one can argue that, from a structural viewpoint, perturbative
renormalizability is a luxury even in Minkowskian quantum field
theories. Of course, it serves as a powerful guiding principle for
selecting physically interesting theories since it ensures that the
predictions of the theory at a certain length scale are independent of
the potential complications at much smaller scales. But it is {\it
not} a consistency check on the mathematical viability of a
theory. Furthermore, in quantum gravity, one is interested precisely
in the physics of the Planck scale; the short-distance complications
are now the issues of primary interest.  Therefore, it seems
inappropriate to elevate perturbative renormalizability to a viability
criterion.

Even if one accepts this premise, however, one is led to ask: Are
there specific reasons that suggest that quantum general relativity
may exist at a non-perturbative level? The answer, I believe, is in
the affirmative.  There are growing indications from a number of
different directions --computer simulations, canonical quantization
and string theory-- that the quantum geometry of space-time would be
quite different from classical geometry [2,4]. Specifically, in this
review we will see that, in a rather wide class of {\it
non-perturbative} quantum gravity theories, the fundamental
excitations of the gravitational field are one dimensional rather than
three; they resemble one-dimensional networks rather than three
dimensional waves. At the Planck scale, geometry has a close
similarity with polymers and the three dimensional continuum arises
only as a ``coarse-grained'' approximation. Perturbative treatments,
on the other hand, assume the validity of a continuum picture at {\it
all} scales. The ultraviolet problems one encounters may simply be a
consequence of the fact that the true microscopic structure of
space-time is captured so poorly in these treatments. Put differently,
if the continuum picture is replaced by a more faithful one, the
``effective dimension'' of space-time could be smaller than four,
whence the theory could have a much better behavior
non-perturbatively.

In this article, I will accept this premise and consider a
non-perturbative quantization of general relativity. Our approach will
differ from the traditional methods of constructing quantum field
theories in a number of ways. First, we will not use any background
metric or connection or indeed any background field. In {\it this}
sense all our constructions will respect the underlying diffeomorphism
invariance of the theory. Second, we will always work in the
continuum.  For actual computations e.g., of the spectra of
interesting operators, we will often introduce ``floating lattices''
or graphs. But these will only serve as computational devices in an
{\it already defined} continuum theory. Third, in the main
constructions, we will refer only to the Planck scale; a
``macroscopic'' scale will not feature in the regularization of
operators or construction of the theory itself%
\footnote{${}^1$}{\ninerm \baselineskip=5pt\noindent Such scales may
be needed to probe the physical meaning of the theory and will then
arise from specific physical problems. For instance, one can seek
semi-classical quantum states that, when coarse-grained on, say, the
weak interaction scale, reproduce classical geometries. The problem
then naturally provides the scale of $\approx 10^{-17}$cm. In the
basic theory itself, these ``macroscopic'' scales do not feature.}.%
 In particular, there will be no cut-offs at a fundamental level.

In the light of the text-book treatments of (flat space) quantum field
theories, these features may seem surprising. Indeed, if we do not
follow one of these standard methods, what then is our strategy?  In
general terms, the idea is to first re-examine the quantization
problem from a somewhat broader perspective. One constructs a suitable
algebra of functions on the classical phase space, promotes it to an
operator algebra and then seeks its representations by operators on a
Hilbert space.  That is, one goes to the ``root'' of the quantization
problem. Due to the presence of an infinite number of degrees of
freedom, this problem is difficult in any field theory. For flat space
field theories, the standard treatments provide strategies to attack
this problem using the powerful machinery associated with the
renormalization group. In the present context, where is there is no
background metric (or any other field), at least at first sight, these
strategies seem not to be as well-suited. What we need is a new
functional calculus that respects the diffeomorphism invariance of the
underlying theory.  Our strategy is to first develop such a calculus
and then use it to construct the Hilbert spaces of states and to
regularize physically interesting operators thereon, directly in the
continuum. The resulting Hilbert spaces and operators are somewhat
unconventional. In particular, the fundamental excitations can not be
identified with gravitons. To particle physicists, this may seem
surprising at first. However, on further thought, one realizes that
this is precisely what one should expect: after all gravitons are
spin-2 excitations on a Minkowskian background and should therefore
arise only as approximate notions in any fully non-perturbative
treatment. There already exist detailed results which indicate how
this can come about [5]. However the relation between our approach and
the standard constructions based on the renormalization group
techniques remains unclear; it would be extremely interesting to
investigate this issue in detail.

The general approach followed in this review is being pursued by a
large number of researchers in about a dozen different groups. I will
not attempt to present a comprehensive or even a systematic survey.
Rather, I will focus only on some of the recent mathematical
developments. These were motivated in part by earlier exploratory work
by a number of colleagues, especially Jacobson, Rovelli and Smolin
[6,7]. For brevity, however, I will forego motivational remarks and
attempt to present the final picture in a concise fashion focusing on
the recent mathematical developments.  (For reviews of the earlier
work, see [8-11].)

\vglue 0.6cm
\leftline{\twelvebf 2. Difficulties and Strategies}
\vglue 0.4cm

The non-perturbative approach I wish to discuss is based on canonical
quantization. The canonical formulation of general relativity was
first obtained in the late fifties and early sixties through a series
of papers by Bergmann, Dirac and Arnowitt, Deser and Misner. In this
formulation, general relativity arises as a dynamical theory of
3-metrics. The framework was therefore named {\it geometrodynamics} by
Wheeler and used as a basis for canonical quantization both by him and
his associates and by Bergmann and his collaborators. The framework of
geometrodynamics has the advantage that classical relativists have a
great deal of geometrical intuition and physical insight into the
nature of the basic variables --3-metrics $g_{ab}$ and extrinsic
curvatures $K_{ab}$.  For these reasons, the framework has played a
dominant role, e.g., in numerical relativity. However, it also has two
important drawbacks.  First, it sets the mathematical treatment of
general relativity quite far from that of theories of other
interactions where the basic dynamical variables are connections
rather than metrics. Second, the equations of the theory are rather
complicated in terms of metrics and extrinsic curvatures; being
non-polynomial, they are difficult to carry over to quantum theory
with a reasonable degree of mathematical precision.

For example, consider the standard Wheeler-DeWitt equation:
$$ 
\big[\sqrt{\textstyle{{G^2\hbar^2\over g}}}
(g^{ab} g^{cd} - {1\over 2} g^{ac}g^{bd})\,\, 
{\delta\over{\delta g_{ac}}} {\delta\over{\delta g_{bd}}}\, - 
\, \sqrt{\textstyle{{g\over{G^2\hbar^2 }}}}\, R(g) \big]\circ 
\Psi (g) = 0\,  ,\eqno(1)
$$
where $g$ is the determinant of the 3-metric $g_{ab}$ and $R$ its
scalar curvature.  As is often emphasized, since the kinetic term
involves products of functional derivatives evaluated at the same
point, it is ill-defined.  However, there are also other, deeper
problems. These arise because, in field theory, the quantum
configuration space --the domain space of wave functions $\Psi$-- is
larger than the classical configuration space. While we can restrict
ourselves to suitably smooth fields in the classical theory, in
quantum field theory, we are forced to allow distributional field
configurations. Indeed, even in the free field theories in Minkowski
space, the Gaussian measure that provides the inner product is
concentrated on genuine distributions. This is the reason why in
quantum theory fields arise as operator-valued distributions.  One
would expect that the situation would be at least as bad in quantum
gravity. If so, even the products of the 3-metrics that appear in
front of the momenta and the meaning of the scalar curvature in the
potential term are obscure. The left hand side of the Wheeler-DeWitt
equation is seriously ill-defined and must be regularized
appropriately.

However, as I just said, the problem of distributional configurations
arises already in the free field theory in Minkowski
space-time. There, we do know how to regularize physically interesting
operators. So, why can we not just apply those techniques in the
present context? The problem is that those techniques are tied to the
presence of a background Minkowski metric. The covariance of the
Gaussian measure, for example, is constructed from the Laplacian
operator on a space-like plane defined by the induced metric and
normal ordering and point-splitting regularizations also make use of
the background geometry. In the present case, we do {\it not} have
background fields at our disposal. We therefore need to find another
avenue. What is needed is a suitable functional calculus --integral
and differential-- that respects the diffeomorphism invariance of the
theory.

What space are we to develop this functional calculus on? Recall first
that, in the canonical approach to diffeomorphism invariant theories
such as general relativity or supergravity, the key mathematical
problem is that of formulating and solving the quantum constraints.
(In Minkowskian quantum field theories, the analogous problem is that
of defining the regularized quantum Hamiltonian operator.) It is
therefore natural to work with variables which, in the classical
theory, simplify the form of the constraints. It turns out that, from
this perspective, connections are better suited than metrics [12].

To see this, recall first that in geometrodynamics we can choose as
our canonical pair, the fields $(E^a_i, G^{-1}K_a^i)$ where $E^a_i$ is
a triad (with density weight one) and $K_a^i$, the extrinsic
curvature.  Here $a$ refers to the tangent space of the 3-manifold and
$i$ is the internal $SO(3)$ --or, $SU(2)$, if we wish to consider
spinorial matter-- index. The triad is the square-root of the metric
in the sense that $E^a_i E^{bi} =: g g^{ab}$, where $g$ is the
determinant of the covariant 3-metric $g_{ab}$, and $K_a^i$ is related
to the extrinsic curvature $K_{ab}$ via: $K_a^i=
(1/\sqrt{g})K_{ab}E^{bi}$.  Let us make a ({\it real}) transformation:
$$ (E^a_i, G^{-1} K_a^i)\,\, \mapsto \,\, (A_a^i:= G^{-1}(\Gamma_a^i -
K_a^i,)\, E^a_i),\eqno(2) $$ 
where $\Gamma_a^i$ is the spin connection determined by the triad.  It
is not difficult to check that this is a canonical transformation on
the real phase space [12,13]. It will be convenient to regard $A_a^i$
as the configuration variable and $E^a_i$ as the conjugate momentum so
that the phase space has the same structure as in the $SU(2)$
Yang-Mills theory. Let us begin by simply writing down the simplest
equations we can, {\it without any reference to a background field},
using these variables. They are: $$ \eqalign{{\cal G}_i& := D_a E^a_i
= 0\cr {\cal V}_b& := E^a_iF_{ab}^i \equiv {\rm Tr}\, E\times B =0\cr
{\cal S} & := \epsilon^{ijk} E^a_i E^b_j F_{abk}\equiv {\rm Tr}\,
E\cdot E\times B =0 \cr}.\eqno(3) $$ These are the ``simplest''
equations in the following sense: Among non-trivial gauge covariant
expressions, the left side of the first equation is the only one which
is at most linear in $E$ and $A$; that of the second is the only one
that is at most linear in $E$ and quadratic in $A$; and, that of the
third is the only one that is at most quadratic in each of $E$ and
$A$. (There is no gauge covariant expression which is linear in $A$
and at most quadratic in $E$.)  Somewhat surprisingly, the first two
equations provide us precisely with the 6 of 7 constraints of general
relativity: the Gauss constraint ${\cal G}_i$ which generates internal
frame rotations and the vector constraint ${\cal V}_b$ that generates
spatial diffeomorphisms [12]. The third equation is almost --but not
quite-- the last (scalar or Hamiltonian) constraint we seek. To see
this, let us translate it in to geometrodynamical variables. It then
reduces to the familiar {\it Euclidean} Hamiltonian constraint [13] $$
R + K^{ab}K_{ab} - K^2 =0, \eqno(3a) $$ rather than the Lorentzian
Hamiltonian constraint $$ R - K^{ab}K_{ab} + K^2 =0, \eqno(3b) $$
which, as one might expect, carries opposite signs in front of terms
which are quadratic in momenta.  (Intuitively, the ``Wick'' rotation
should map the momentum to $i$ times the momentum.) The Gauss and the
vector constraints are insensitive to the signature.

To get the Lorentzian theory, the most straightforward avenue is to
consider {\it complex} connections ${}^{c}\!A_a{}^i := \Gamma_a^i - i
K_a^i$. This avenue has been pursued vigorously in the literature
[8-11]. In this article, however, I will present another possibility
which has arisen from a recent idea of Thomas Thiemann's and which may
allow one to pass from the solutions to the Euclidean quantum
constraints to the Lorentzian ones via a generalized Wick
transformation. This strategy will be discussed in Section 5. For the
moment, let us note only that $SU(2)$ connections can be regarded as
the configuration variables also in general relativity. Thus, the
classical configuration space is $\ag$, the space of suitably smooth
connections modulo gauge transformations. Our task is to find the
corresponding quantum configuration space and to develop functional
calculus on it.

I will conclude this discussion with a curiosity: The Euclidean
signature seems to be easier to handle in other approaches as well, in
particular, Connes' non-commutative geometry and string theory. In the
Connes' ``geometrization'' of the standard model, one is naturally led
to work with the Euclidean signature and the extension of the
framework to the Lorentzian regime is still to be worked out. In
string theory, while one can allow the target space geometry to be
Lorentzian, in all the discussions that I am aware of, the world sheet
is kept Euclidean. In perturbation theory off Minkowskian backgrounds,
as far as the S-matrix theory is concerned, this seems like an obvious
extension of the strategy one adopts in field theory.  However, in
non-perturbative contexts, such as the discussion of mirror symmetry,
restriction to Riemannian world-sheets appears artificial from a
physical perspective. As in the above discussion of the Hamiltonian
constraint, in both these examples, one needs an appropriate,
generalized Wick transform.

\vglue 0.6cm
\leftline{\twelvebf 3. Tools}
\vglue 0.4cm

The classical configuration space $\ag$ is the quotient of the space
$\a$ of suitably smooth connections by the group $\g$ of local gauge
transformations. $\ag$ can be endowed with topology in a standard way,
using a Sobolev norm on the space $\a$. The key question now is: What
is the {\it quantum} configuration space? More precisely, in the
connection representation, what is the domain space of quantum wave
functions?  In quantum mechanics --i.e., in the quantum theory of
systems with a finite number of degrees of freedom-- the classical and
the quantum configuration spaces agree. In field theory, as noted in
Section 1, the quantum configuration space is a substantial
enlargement of the classical one; in scalar field theories, for
example, although the classical configurations are smooth (say $C^2$)
functions on a $t={\rm const}$ slice, the quantum configuration space
consists of all tempered distributions. Furthermore, this enlargement
is not a mere technicality: the set of smooth configurations is of
zero measure with respect to the Gaussian measure that determines the
inner product! The regularization problems of quantum field theory can
be traced back to this fact.

The text-book treatments for the construction of quantum configuration
spaces are not directly applicable in the present case because they
are geared to the case in which classical configuration spaces are
linear. Nonetheless, a natural strategy {\it is} available
[14]. Recall first that the Wilson-loop functions, $T_{\alpha}(A) :=
\textstyle{1\over 2} {\rm Tr} {\cal P} \, \exp \oint A.dl$, form a
natural (over)complete set of functions on $\ag$ in the sense that
they suffice to separate the points of $\ag$. (Trace is taken in the
fundamental representation.) They are thus natural candidates for our
configuration variables%
\footnote{${}^2$}{\ninerm
\baselineskip=5pt\noindent All loops will be based at some arbitrarily
chosen but fixed point on $\Sigma$. For technical reasons, we will
restrict ourselves to piecewise analytic loops. Some of the key
results have been extended to the smooth category by Baez and Sawin
[15].}. 
It is straightforward to construct a $C^\star$-algebra
generated by these $T_\alpha$. It is called the {\it holonomy algebra}
and denoted by $\HA$.  The first step in the quantization procedure is
to develop the representation theory of $\HA$.  A natural candidate
for the quantum configuration space will arise from this theory.

Since elements of this algebra are all configuration variables, the
algebra is Abelian and it is equipped with an identity, the Wilson
loop function associated with the trivial (point) loop. Now, the
representation theory of Abelian $C^\star$-algebras with identity has
been developed in detail by Gel'fand and Naimark. One of their basic
results is that any such $C^\star$-algebra is naturally isomorphic
with the $C^\star$-algebra of {\it all} continuous functions on a
compact, Hausdorff space, called the spectrum of the algebra. Thus,
every element $a$ of the algebra is canonically represented by a
concrete function $\check{a}$ on the spectrum. In our case, the
algebra $\HA$ is smaller than the algebra of all bounded continuous
functions on $\ag$ and the spectrum turns out to be larger than
$\ag$. However, since the elements of $\HA$ suffice to separate points
of $\ag$, it follows that $\ag$ is densely embedded in the spectrum
[16]. Thus, we can regard the spectrum as a completion of $\ag$. To
emphasize this point, we will denote it by $\agb$.

Since $\agb$ is a compact Hausdorff space, using the Riesz
representation theorem, one can conclude that every cyclic
representation of $\HA$ by bounded operators on a Hilbert space is of
the following type: The Hilbert space is $L^2(\agb, d\mu)$ for some
measure $\mu$ and the operators $\hat{T}_\alpha$ act via:
$$
(\hat{T}_\alpha \circ \Psi )(\Ab) = \check{T}_\alpha(\Ab) 
\Psi (\Ab)\, ,\eqno(4)
$$
where the function $\check{T}$ on $\agb$ is the Gel'fand transform of
$T_\alpha$. Thus, $\agb$ serves as an universal domain space for
quantum states; it is therefore the {\it quantum configuration space}.  

The details of this construction {\it will not be needed} in what
follows.  One just needs to note that a natural completion $\agb$ of
$\ag$ will serve as the quantum configuration space and that, even
though $\ag$ is a rather complicated, infinite dimensional space,
$\agb$ is compact.

Our task then is to develop integral and differential calculus on
$\agb$: measures and integration theory are needed to specify the
inner-product and differential operators are needed to define
observables. The task has been carried out in a series of papers
[17-24].  In essence, this development was possible because there are
three distinct ways of characterizing $\agb$, each illuminating a
specific aspect of its structure: first, as the Gel'fand spectrum of
the holonomy algebra [14]; second, as the space of homomorphisms from
the so-called hoop group%
\footnote{${}^3$}{\ninerm
\baselineskip=5pt\noindent The hoop group ${\cal HG}$ of the
3-manifold $\Sigma$ is defined as follows. Consider the space of
piecewise analytic, closed loops in $\Sigma$ which begin and end at
the fixed point $x$. Regard two loops as equivalent if the holonomy of
any (smooth) $SU(2)$ connection around one equals that around the
other.  The space of equivalence classes then has a natural group
structure, groups corresponding to different base points being
isomorphic. This group is ${\cal HG}$. While the definition refers to
a specific gauge group, the resulting ${\cal HG}$ turns out to be
largely independent of this choice: there are only two hoop groups;
one Abelian (associated with Abelian gauge groups) and the other
non-Abelian [17].}  
to the gauge group $SU(2)$ [17]; and third, as a projective limit of
configuration spaces associated with graphs in $\Sigma$ [19,20]. As a
result, $\agb$ has structure which is much richer than what one might
expect at first.

That integral calculus can be developed [17-20] is perhaps not
surprising: after all, $\agb$ is a compact, Hausdorff space and
therefore admits regular measures. The development of a differential
calculus, on the other hand, seems hopeless at first since Gel'fand
spectra are only topological spaces; the Gel'fand theory does {\it
not} endow them with a differential structure.  In particular, it
seems difficult to convert $\agb$ into an infinite dimensional
manifold. However, {\it because our $C^\star$-algebra is special}
--the holonomies have a natural geometrical meaning-- $\agb$ can be
regarded as the projective limit of finite dimensional manifolds (the
third characterization above). This enables one to introduce the
notion of smooth functions on $\agb$ and use it to introduce forms,
vector fields, and a number of differential and integral operators
{\it without any reference to a background metric} on the underlying
3-manifold $\Sigma$ [21]. We will see some illustrations of how this
is achieved in the next section.

To proceed further in a concrete fashion, one has to choose a specific
measure on $\agb$ and work in the representation of the holonomy
algebra it provides. The algebra itself imposes no restriction on this
choice; for {\it any} regular measure $\mu$, the Wilson loop operators
$\hat{T}_\alpha$ are bounded {\it self-adjoint} operators on the
resulting Hilbert space $H \equiv L^2(\agb , d\mu)$. It is the
requirement that the {\it momentum} operators be self-adjoint that
restricts the measure. (A similar situation occurs already in
non-relativistic quantum mechanics: while the position operator
$\hat{X}$ is self-adjoint on $L^2(R, fdx)$ for any (regular) function
$f$, the momentum operator $\hat{P}\equiv -i\hbar d/dx$ is self
adjoint only if $f$ is a constant. Thus, it is the self-adjointness of
the explicitly defined momentum operators that singles out the
Lebesgue measure $dx$.) In our case, the momenta are essentially the
electric fields. However, electric fields are not gauge invariant and
to obtain an operator which descends to $\ag$ --i.e., has a meaningful
action on the space of gauge invariant wave functions-- further work
is needed%
\footnote{${}^4$}{\ninerm \baselineskip=5pt\noindent The simplest 
strategy is to use 2-dimensional strips $S$, parametrized by $(\sigma,
\tau)$, which are foliated by closed loops $\tau = {\rm const.}$, and
consider the associated ``momentum variables'' $P_S(A, E) := \int
dS^{ab}\epsilon_{abc}\- ({\rm Tr} E^c(\sigma, \tau) U(\sigma, \tau))$,
where $U$ is the holonomy of $A$ around the loop $\tau = {\rm const}$,
evaluated at the point $(\sigma, \tau)$ of $S$. This is a gauge
invariant function on the phase space, linear in the momentum $E$, and
depends only on the foliation of $S$ (i.e., not on the details of the
parameterization).  Since there are no background fields, the Poisson
algebra of configuration and momentum variables, $T_\alpha$ and $P_S$,
can be expressed in an elegant fashion, using only the topological
properties of loops and strips in 3-dimensions [25].}. %
When this is done, one obtains momentum operators that can act on
suitably regular functions $\Psi(\Ab)$ on $\agb$. The requirement that
these momentum operators be self-adjoint then picks out a measure
$\mu_o$ on $\agb$ [23].

This measure has four important properties [17-19]: i) it is
normalized; the total $\mu_o$ measure of $\agb$ is $1$; ii) it is
faithful; the integral of every non-negative, continuous function on
$\agb$ is non-negative and vanishes if and only if the function is
identically zero; iii) $\mu_o$ is invariant under the (induced action
on $\agb$ of the) diffeomorphism group on the underlying 3-manifold
$\Sigma$; and, iv) $\mu_o$ is concentrated on genuinely generalized
connections; the $\mu_o$-measure of the classical configuration space
$\ag$ is zero! The last result is the analog of the standard situation
in flat space quantum field theory referred to above; the classical
configuration space is topologically dense in the quantum
configuration space but measure theoretically sparse.

We can use this measure to construct a Hilbert space $H_o := L^2(\agb,
d\mu_o)$. This is the space of kinematic states of quantum gravity. It
is ``kinematic'' because we are yet to impose quantum constraints to
select physical states which have dynamical information; it is the
quantum analog of the {\it full} phase space of general relativity.
To solve quantum dynamics, we have to first regularize quantum
constraints and express them as well-defined operators on $H_o$ and
then isolate the kernel of these operators. We will turn to this task
in Section 5.

We will conclude this section by pointing out two properties of $H_o$
which will play an important role in the subsequent discussion.

First, $H_o$ admits [24] an interesting orthonormal basis, obtained by
a generalization of Penrose's spin networks [27] (which, although
introduced in quite a different context, were also motivated by
quantum gravity considerations. See also [26].). A penrose network is
a closed graph, each of whose vertices is trivalent (i.e., has three
edges), with an assignment of representation of SU(2) (i.e., of a
half-integral number $j$) to each of its edges, such that, if $j_1,
j_2, j_3$ label edges incident at any one vertex, we must have $|j_1
-j_2| \le j_3 \le j_1+j_2$ (and permutations thereof). Now, each such
network $N$ defines a function $\Psi_N$ on $\agb$ as follows. First,
one can show [18,21] that every (generalized) connection $\Ab$ assigns
to each edge an element of $SU(2)$, the holonomy.  Using this
assignment, we can associate a matrix with each edge in the Penrose
network. The constraint on the choice of representations ensures that
all the resulting matrices can be consistently contracted to produce a
number, the value of $\Psi_N$ on the chosen point $\Ab$ of
$\agb$. Remarkably, these states are orthonormal in $H_o$. However,
they are not complete.  To obtain a complete set, we have to allow
higher valent networks --i.e., the ones which have an arbitrary (but
finite) number of edges at any one vertex-- assign to each edge a
representation of $SU(2)$ {\it and} to each vertex a suitable
contractor (or, an inter-twiner).  The resulting space of states can
then be used to obtain an orthonormal decomposition of $H_o$ into {\it
finite} dimensional subspaces, each labelled by the network (embedded
in $\Sigma$) and the assignment of representations [24,23]. (The
dimensionality of each subspace is given by the number of independent
contractors and one can introduce an orthonormal basis in each
subspace using the Schmidt procedure.)

The second property of $H_o$ is that it admits an interesting dense
subspace. Fix a closed graph $\gamma$ (i.e., a collection of vertices
joined by analytic edges such that every vertex has at least
two edges) in $\Sigma$. Every (generalized) connection $\Ab$ 
associates to each edge $e_i$ of $\gamma$ an element $g_i$ of $SU(2)$.
Thus, there is a natural projection from $\agb$ to $[SU(2)]^n$, where
$n$ is the total number of edges in the graph.  Using this projection,
we can pull-back to $\agb$ functions on $[SU(2)]^n$. These pull-backs,
$$ 
\Psi_\gamma (\Ab) := \psi (g_1(\Ab),\- ....\-, g_n(\Ab))\, ,\eqno(5)
$$
for some smooth function $\psi$ on $[SU(2)]^n$ are called {\it
cylindrical functions} on $\agb$. Note that the functions
$\Psi_\gamma$ only know about what the connection $\Ab$ is doing at
points of $\Sigma$ which lie in the graph $\gamma$. Yet, as we vary
the graph and consider more and more vertices and edges, we obtain a
bigger and bigger collection of functions on $\agb$. They are all
square-integrable with respect to $\mu_o$ and, furthermore, span a {\it
dense} subspace of $H_o$. This space $\Cyl (\agb)$ of all cylindrical
functions plays an important role in calculations, rather analogous to
that played by the space $C^\infty_o(R)$ of smooth functions of
compact support in quantum mechanics on a line. Just as we often first
define operators on $C^\infty_o(R)$ and then extend them to (other,
better suited dense subspaces of) $L^2(R, dx)$, in the present
context, we will often define operators first on $\Cyl(\agb)$ and then
extend them.

These two properties of $H_o$ provide considerable intuition about the
nature of quantum states that we have been led to consider. Since an
orthonormal basis is provided by networks, it follows that the
``elementary excitations'' are {\it 1-dimensional} rather than
three. They are ``loopy'' rather than ``wavy''. A typical excited
state looks like a polymer rather than a smooth undulation on flat
space. Just as a polymer in a sufficiently complex configuration
behaves as if it were a 3-dimensional entity, we will see that the
1-dimensional excitations, if packed densely and superposed
coherently, can approximate a 3-dimensional continuum geometry very
well. The quantum states we have encountered here are qualitatively
different from the ones we come across in Minkowskian quantum field
theories. The main reason is the underlying diffeomorphism invariance:
In absence of a background geometry it is not possible to introduce
the familiar Gaussian measures and associated Fock spaces.

\vglue 0.6cm
\leftline{\twelvebf 4. Quantum Geometry}
\vglue 0.4cm

The framework constructed in Section 3 provides us with sufficient
tools to explore the nature of quantum geometry and, as we will see,
the results are quite surprising.

Let us first recall that, in classical general relativity, geometrical
observables can be regarded as functions on the phase space. Let us,
for example, fix a smooth 2-surface $S_o$ or a 3-dimensional region
$R_o$ in the 3-manifold $\Sigma$. Then, given any triad field $E^a_i$,
we can assign to $S_o$ an area, $A(S_o)$, and to $R_o$ a volume,
$V(R_o)$:
$$
A(S_o) := \int_{S_o}|E^3_i\, E^{3i}|^{1\over 2} d^2x; \quad {\rm and}
\quad V(R_o) := \int_{R_o} |{\rm det} E|^{1\over 2} d^3x\,  , 
\eqno(6)
$$
where, to simplify the presentation, we have chosen coordinates so
that $S_o$ is given by $X_3 = {\rm const}$. Thus, $A(S_o)$ and
$V(R_o)$ can be regarded as functions on the {\it full} classical
phase space (which happen to depend only on the triad, i.e. happen to
be independent of the connection). Our problem now is to promote these
functions to operators on the Hilbert space $H_o$ --the quantum analog
of the full phase space-- and study their properties.

At first sight, the challenge seems unsurmountable. The first obstacle
is that even the classical expressions are {\it non-polynomial} in the
triads.  Second, our wave functions $\Psi(\Ab)$ have support on
generalized connections which have a ``distributional character.''
Hence, the task of regularizing formal expressions like
$$ |- ({\delta\over\delta A_3^i(x)}) ({\delta\over\delta
  A_3^i(x)})|^{1\over 2}$$ 
seems formidable. In addition, we do not have a background metric to
simplify this task; we thus have the constraint that the final,
regularized operators should not depend on any background fields. It
turns out, however, that, using the functional calculus on $\agb$, all
these difficulties can be overcome [28, 23]. (For an approach based on
the loop representation, see [29] and, on lattice regularization, see
[30]. I should add however that I do not yet know the precise relation
to these approaches.)

One begins with the classical expressions, introduces a chart and
point splits the triad fields using regulators, i.e., 2-point
functions $f_\epsilon(x,y)$ which tend to $\delta^3(x,y)$ in the limit
as $\epsilon$ tends to zero. One then notices that the resulting
expressions can be promoted to operators which have a well-defined
action on cylindrical functions on $\agb$. One evaluates this action
and, in the final expression, takes the limit $\epsilon \mapsto 0$.
The limits are well-defined and yield operators which carry no memory
of the background chart or regulators used in the procedure. They map
cylindrical functions based on a graph $\gamma$ to cylindrical
functions based on the {\it same} graph. Let us fix a graph with n
edges. Then, given a cylindrical function $\Psi_\gamma(\Ab)$ which is
the pull-back of a function $\psi (g_1, ..., g_n)$ of $[SU(2)]^n$, the
area operator yields:
$$
\hat{A}(S_o)\circ \Psi (\Ab) := \ell_P^2\, \sum_\alpha\, 
|\sum_{I_\alpha J_\alpha}\, K(I_\alpha, J_\alpha) X_{I_\alpha}^i
X_{J_\alpha}^i|^{1\over 2}\, \circ \psi(g_1, ..., g_n)\, . \eqno(7)
$$
Here $\ell_P$ is the Planck length; $\alpha$ ranges over the vertices
of the graph $\gamma$ which intersect $S_o$; $I_\alpha$ and $J_\alpha$
denote edges passing through the vertex $\alpha$; $K(I_\alpha,
J_\alpha)$ takes values $0, \pm 1$ depending on the orientation of the
two edges relative to $S_o$; and, $X_{I}^i$ is the $i$-th right (left)
invariant vector field on the $I_\alpha$-th copy of the gauge group if
the edge $I_\alpha$ is outgoing (incoming) at the vertex
$\alpha$. (Note that since there are three right (left) invariant
vector fields on $SU(2)$, the index $i$ runs over $1,2,3$.) Each
vector field $X$ acts on only one of the arguments in the wave
function, i.e., on the copy of the gauge group picked by the
associated edge.  Similarly, the volume operator is given by:
$$
\hat{V}(R_o)\circ \Psi(\Ab) := \ell_P^3\, \sum_\alpha \,\, 
|\sum_{I_\alpha, J_\alpha, K_\alpha} i\, \epsilon^{ijk}\, 
\epsilon(I_\alpha, J_\alpha, K_\alpha)\, 
X^i_{I_\alpha}X^j_{J_\alpha} X^k_{K_\alpha}|^{1\over 2}\, 
\circ \psi(g_1, ..., g_n)\, , \eqno(8)
$$
where the first sum now is over vertices which lie in the region R and
$\epsilon(I_\alpha, J_\alpha, K_\alpha)$ is $0$ if the three edges are
linearly dependent at the vertex $\alpha$ and otherwise $\pm 1$
depending on the orientation they define.  Thus, the two operators are
well-defined on the space $\Cyl(\agb)$ of cylindrical functions on
$\agb$. Recall, however, that this space is dense in $H_o$. Using this
fact and the properties of right and left invariant vector fields on
$SU(2)$, one can show that they admit unique self-adjoint extensions
on $H_o$.

For our purposes here, {\it the details of these operators are not
important}. It would be sufficient to note just that closed form
expressions of the fully regulated operators are available and they
involve the actions of right and left invariant vector fields on the
appropriate copies of $SU(2)$. Because these actions are completely
understood, in practice, it is rather straightforward to compute how
these operators act on cylindrical states associated with specific
graphs.

A number of remarks are in order.

1. Given that the expressions of the classical observables are already
rather involved, why are the expressions of the quantum operators
relatively simple? The main reason is that the requirement of
diffeomorphism invariance --i.e., absence of background fields--
severely restricts the possible operators. Consider, for example, the
area operator. Given a graph and a surface, the obvious diffeomorphism
invariant notion is that of an intersection of the graph with the
surface. Not surprisingly, the expression of the operator is a sum
over intersections. At each intersection, the graph has a number of
edges. So, all that the operator can do is to act on the copies of
groups associated with these edges. That is, thanks to diffeomorphism
invariance, the action of operators is reduced to simple algebraic
operations in the representation theory of $SU(2)$!

2. Since the two operators are self-adjoint, one can compute their
spectra.  {\it They are purely discrete!}  Each of the Penrose spin
network state is an eigenstate of {\it both} the operators. The area
operators reduce to a sum of $SU(2)$-Laplacians, each of which has the
familiar eigenvalues $\sqrt{j(j+1)}$. The eigenvalue of the volume
operator on these states is, however, zero [30, 28]. Thus, higher
valent spin networks are essential. Both operators again leave the
(finite dimensional) Hilbert space associated with any of these
networks invariant. They can therefore be diagonalized separately on
each network.  (Already in the 4-valent case, there exist states with
non-zero volume.) This overall picture shows that {\it the microscopic
geometry at Planck scale is very different from what the
continuum picture suggests}. As we argued in the Introduction, this
basic fact may well be the essential reason why the perturbation
theory fails.

3. The geometrical observables can be defined on the full phase space
of the classical theory; one need not restrict oneself to the
constraint surface. Similarly, in the quantum theory, the analogous
operators have been defined on the kinematic Hilbert space $H_o$,
before imposing quantum constraints. The results are thus robust; they
are not sensitive to the details of quantum dynamics. For example,
they will continue to hold also in supergravity, and more generally,
in any theory in which the triads and connections feature as a basic
canonical pair.

4. There is, however, a subtlety. To see this, recall a basic
difference between quantum mechanics and quantum field theory.  In
quantum mechanics, kinematics and dynamics are largely decoupled in
the sense the same representation of the canonical commutation
relations supports all interesting Hamiltonians. In quantum field
theory, on the other hand, kinematics and dynamics are always weakly
coupled: there are infinitely many {\it inequivalent} representations
of the canonical commutation relations and, in general, different
Hamiltonians are meaningful on different representations. In the
present case, then, the key question is whether the quantum
constraints of a given theory can be meaningfully imposed on
$H_o$. Results to date indicate (but do not yet prove) that this is
likely to be the case for constraints of general relativity. My
general expectation is that this will also be the case for a class of
theories such as supergravity which are ``near general relativity'' as
far as geometry is concerned.  What we have here is a glimpse into the
nature of quantum geometry underlying {\it this} class of quantum
gravity theories.

5. Note that $\hat{A}(S)$ and $\hat{V}(R)$ have been defined for {\it
any} surface $S$ and {\it any} region R in $\Sigma$ and will therefore
not be Dirac observables; they are not expected to commute with the
quantum constraints. To obtain Dirac observables, one would have to
specify $S$ and $R$ {\it intrinsically}. (For example, if we had only
the Gauss and the diffeomorphism constraints, the total volume of
$\Sigma$ would be a Dirac observable.) A natural strategy is to
specify $S$ and $R$ using matter fields (see, e.g.,[31]). In view of
the Hamiltonian constraint, the problem of providing an explicit
specification of this type is extremely difficult. However, {\it if} a
surface $S$ or a region $R$ were specified in this manner, Eqs (6)
and (7) will provide the expressions of the associated area and volume
operators.

6. Note that the expressions of the operators do not involve any free
re-normalization constants; they are completely unambiguous. For the
two operators considered here, this is just a fact of
calculations. However, there exist heuristic arguments that indicate
that, in a more general context, if the end result of such a
regularization procedure is an operator which has no background
dependence, then these final operators should be free of arbitrary
constants [10]. This reasoning is borne out in the construction of
another geometric operator, associated with the classical observable
$E(\omega):= \int d^3x (E^a_i E^{bi} \omega_a \omega_b)^{1\over 2}$,
where $\omega$ is a smooth 1-form of compact support on $\Sigma$.  (As
in the case of area and volume, a square-root is essential to obtain a
density of weight one which one can then integrate without reference
to any background metric or volume element. Note incidentally that,
since $\omega_a$ is arbitrary, classically, the functional $E(\omega)$
has the full information about the metric.) It is not known if there
are other well-defined geometrical operators, e.g. one which carries
information about the length functional directly.

\vglue 0.6cm
\leftline{\twelvebf 5. Quantum Einstein Equations}
\vglue 0.4cm

We now turn to quantum dynamics. In the canonical approach, this is
captured in the quantum constraints. Our task then is the following:
i) Write the regulated constraints as operators on the kinematic
Hilbert space $H_o$; ii) Check for anomalies; iii) If there are none,
``solve'' the quantum constraints; and, iv) On the space of solutions,
introduce an appropriate Hilbert space structure. If all this can be
achieved, one would have a coherent mathematical framework; one would
say that quantum general relativity does exist non-perturbatively. One
would, of course, still have to devise suitable approximation schemes
to extract physical predictions.

There is however a key difficulty, of quite a general sort, associated
with quantum constraints that needs to be resolved before we proceed
further. To illustrate it, let us consider the example of a free
relativistic particle in Minkowski space. Then, the classical
configuration space is the 4-dimensional Minkowski space-time, the
phase space is the cotangent bundle over it and there is a single {\it
dynamical} constraint $C(x,p) := \eta^{ab}p_ap_b + m^2 =0$.  In
quantum theory then, we can take $H_o := L^2(R^4, d^4x)$ to be the
fiducial, {\it kinematical} Hilbert space. The constraint can be
promoted to a self-adjoint operator $\hat{C}:= -(\eta^{ab}
\partial_a \partial_b - \mu^2)$ on $H_o$ (where $\mu = m/\hbar$). 
Thus solutions $\Psi(x)$ to the quantum constraint are easily
obtained: they solve the Klein-Gordon equation. The problem is that
none of the non-zero solutions lies in the kinematical Hilbert space
$H_o$. (This is obvious in the momentum representation where the
solutions $\psi(p)$ are distributions $\psi(p) = g(p) \delta (p.p +
m^2)$ for some well-behaved function $g(p)$ and therefore not
square-integrable in the $L^2$-norm.) This is a rather general
problem; it is in the exceptional case when the group generated by the
constraint functions is compact that the solutions $\Psi$ ---i.e.,
physical states-- belong to $H_o$. Thus, in the generic situation, we
face two problems: What is the ``home'' of the physical quantum
states? Once the home has been found and space of suitable solutions
isolated, how is one to introduce an inner product on this space?

Fortunately, there exists [32] a general approach to both these
problem, based on the idea of ``averaging over'' the orbit of the
group generated by the constraints. In the present case, one first
considers the 1-parameter group of unitary transformations,
$U(\lambda):= \exp\, i\lambda \hat{C}$, generated by the constraint
$\hat{C}$. The idea is to construct solutions by averaging suitable
elements of $H_o$ over the 1-parameter group generated by
$U(\lambda)$. For the result to be manageable, however, we have to
restrict the initial element to a ``nice'' (dense) subspace ${\cal S}$
of $H_o$. Let us choose this ${\cal S}$ to be the space of smooth test
functions of compact support.  Then, given any test field $f$, one can
show that $\Psi_f :=\int d\lambda U(\lambda)\circ f$ exists as a
well-defined distribution over ${\cal S}$ and solves the quantum
constraint. Thus, the natural home for these solutions is ${\cal
S}^\star$, the topological dual to ${\cal S}$ we began
with. Furthermore, we can now naturally define an inner product:
$(\Psi_f, \Psi_h) := \Psi_f(h) \equiv \int d^4x \bar{\Psi}_f(x) h(x)$.
It turns out that this is the correct inner product. Thus, in this
example, the group averaging procedure provides an answer to both our
questions. (As far as I know, however, there are no general theorems
to ensure a priori that the method is so directly applicable in more
general situations. The practical strategy is to apply it to any given
case and see if the final inner-product is well-defined, i.e.,
positive definite.)

The procedure {\it is} applicable to the vector (or the
diffeomorphism) constraint of general relativity [23]. One first
promotes the formal ``exponentiated forms'' of diffeomorphism
constraints to well-defined, unitary operators on the kinematical
Hilbert space $H_o$.  (These operators are unitary because the measure
$\mu_o$ is diffeomorphism invariant.) In the second step, one checks
for anomalies.  In various lattice regularizations, these were
encountered essentially because the lattice regularization manifestly
breaks the diffeomorphism invariance. Here, we are working directly in
the continuum, have a well-defined Hilbert space and unitary operators
$U_{\vec N}(\lambda)$ thereon, for each analytic vector field ${\vec N}$
on $\Sigma$. We can directly check their algebra. {\it One finds that
there are no anomalies.} One can now proceed to the last two steps:
solving the constraint and introducing the inner product on the space
of physical states. Here we apply the group averaging technique.  As
one might expect, the role of the dense sub-space ${\cal S}$ is played
by the space of cylindrical functions $\Cyl(\agb)$ and the physical
states lie in its topological dual. They are not normalizable with
respect to $H_o$. Nonetheless, using the procedure outlined above, one
can endow them with an inner product and obtain a Hilbert space $H_d$
of solutions to the diffeomorphism constraints%
\footnote{${}^5$}{\ninerm \baselineskip=5pt\noindent In the analytic 
category used in this paper, we have been able to obtain an infinite
dimensional family of solutions but it is not clear if this class
exhausts all physically interesting situations. This problem would
disappear if we could extend the entire discussion to the physically
better suited smooth category along the lines initiated by Baez and
Sawin [15]. For details, see [23].}. %
Roughly, states in $H_d$ can be labelled by equivalence classes of
spin networks, where two are regarded as equivalent if they are
related by the action of the diffeomorphism group. This provides a
precise formulation of the interplay between knot theory and general
relativity that was anticipated by Rovelli and Smolin already in [7]
and also brings out a number of subtleties. Finally, the inner-product
on $H_d$ automatically incorporates the ``reality conditions'' [8,33]
correctly.  More precisely, the projection to $H_d$ of operators on
$H_o$ which are self-adjoint and commute with the diffeomorphism
constraints are guaranteed to be self-adjoint on $H_d$.

The Gauss constraint is already taken care of since we are working on
the space $\agb$ of gauge equivalent (generalized) connections.
Alternatively, as indicated in [23], one could first begin with the
space $\overline{\cal A}$ of (generalized) connections, show that there
are no anomalies, impose the Gauss constraint using group averaging,
and show that the resulting Hilbert space of states is precisely the space
$H_o$ that we began with in this paper.

Finally, we come to the difficult Hamiltonian constraint, the analog
of the Wheeler-DeWitt equation. There has been considerable work on
solutions to this constraint (see, e.g., [34-37]) and several
fascinating results have been obtained. For example, some of the
solutions have been identified with well-known knot
invariants. However, most of this work is of a rather formal nature;
for example, it is generally unclear if the underlying theory is
Euclidean or Lorentzian. In keeping with the general spirit of this
review, here, I will set these results aside and consider instead a more
recent development which focuses more on the structural issues rather
than on specific solutions.

Let us begin with the third constraint in Eq (3), even though it
refers to the Euclidean --rather than the Lorentzian-- theory.  The
task is to define the corresponding quantum constraint operators on
the Hilbert space $H_d$ of solutions to the diffeomorphism
constraints. Using a key idea due to Rovelli and Smolin [38], a
regulated form of this operator has been constructed and it has been
shown that, in the appropriate sense, the constraint algebra closes
without anomalies [39]. While this progress is notable, I should point
out that, in contrast to the regularization of geometrical operators
or the imposition of the diffeomorphism constraint, these calculations
should be regarded as preliminary explorations. Specifically, there is
considerable freedom at an intermediate step and no compelling reason
to adopt the specific prescription used there. Furthermore, properties
of the resulting constraint operator have not been investigated in
detail.  Even heuristically, it does not appear to be self-adjoint. On
the one hand, as Kucha\v{r} [40] has pointed out, this would
consistent with the closure of the constraint algebra since in finite
dimensional models with an analogous algebra, the constraints would
not close if the analog of the Hamiltonian constraint were
self-adjoint (although this possibility is not ruled out in infinite
dimensions due to subtleties associated with regularization.) However,
non self-adjointness would also make it impossible to use the group
averaging method at least directly.  In spite of these important
limitations, the calculation does represent a first serious attempt to
regularize the Hamiltonian constraint rigorously and therefore holds
considerable promise.  In geometrodynamics, for example, a comparable
stage is yet to be reached with respect to the Wheeler-DeWitt
equation.

There is, however, a much more serious difficulty: The classical
constraint we began with refers to the {\it Euclidean} signature
while, physically, we need to solve the Lorentzian constraint. Thus,
we need a generalized Wick transformation which will map the solutions
to the quantum Euclidean constraint to those of the Lorentzian.  At
first, the problem looks hopelessly difficult. However, recently,
Thiemann [41] has suggested a strategy in the context of a ``coherent
state transform'' that will map complex-valued functions of real
$SU(2)$ connections to holomorphic functions of an $SL(2,C)$
connection. The same strategy can be adopted in a purely real
formulation used in this report (and also extended to allow for the
presence of matter sources) [42]. To see how this works, let us return
to the classical theory, where the phase space, canonical variables
and constraints are all real in both signatures. The idea is to define
an automorphism $W$ on the algebra of complex-valued functions on the
phase space --i.e., a map which preserves linear combinations,
products and Poisson brackets of functions-- which {\it maps the
Euclidean Hamiltonian constraint functional (3a) to the Lorentzian one
(3b)}. $W$ is generated by a phase-space function $T$ as follows:
$$
W\circ f := f + \{T, f\} + \textstyle{1\over 2!} \{T,\{T, f\}\} + ...
\,\, \equiv \sum_{i=0}^\infty \, \textstyle {1\over n!}\{T, f\}_n\, , 
\eqno(9)
$$
where $\{T, f\}$ is the Poisson bracket between $T$ and $f$ and $\{T,
f\}_n$, the repeated Poisson bracket of $n$ $T$-factors with $f$ and
the generating function $T$ is given by:
$$
T := \textstyle{i\pi\over 2}\int_\Sigma\, d^3x\, K_a^i E^a_i\,\,
\equiv {\textstyle{i\pi\over 2}}\{V,  H_E\}\, .\eqno(10)
$$
Here, $V$ denotes the total volume functional and $H_E := \int d^3x
g^{-{1\over 2}}\,{\cal S}$, where, as before, ${\cal S}$ is the
Euclidean scalar or Hamiltonian constraint (see Eq (1.3). Note that
$T$ is just the integral of the trace of the extrinsic curvature over
$\Sigma$.) It is straightforward to verify that $W\circ {\cal S} =
{\cal S}_L$.  Hence, if we could promote $T$ to a quantum operator, we
would have the generalized Wick transform $\hat{W}:=\exp
-\hat{T}/i\hbar$ in the quantum theory which would map the solutions
to the Euclidean Hamiltonian constraint to those of the Lorentzian: We
would have
$$
\hat{\cal S}\circ \Psi = 0\quad \Longrightarrow \quad \hat{\cal S}_L
\circ (\hat W\circ\Psi) = 0\, , \eqno(11)
$$
$\hat{\cal S}_L$ being the Lorentzian Hamiltonian constraint. That is,
while we did not have, a priori, a way of defining the Lorentzian
Hamiltonian constraint operator, since classically $W$ maps the
Euclidean Hamiltonian constraint functional to the Lorentzian one, we
can simply {\it define} the Lorentzian quantum constraint $\hat{\cal
S}_L$ by $\hat{\cal S}_L = \hat W\circ \hat {\cal S}\circ \hat W^{-1}$.
This is an attractive strategy especially since, as Eq (10) shows,
$\hat{T}$ could be constructed from the total volume operator (on
which we have full control) and the integrated Euclidean Hamiltonian
constraint.

To summarize, {\it if} the Euclidean quantum constraint $\hat{\cal S}$
can be regularized in a more satisfactory fashion (so as to be free of
the drawbacks of the present scheme discussed above) and the operator
$\hat{T}$ defined and shown to have certain properties on $H_d$, we
would conclude that {\it all} quantum constraints of general
relativity can be imposed consistently, i.e., that quantum general
relativity is consistent at a non-perturbative level.  Note that it is
not essential to exhibit the general solution to the Hamiltonian
constraint; indeed, even in the classical case, we do not have a
general solution to the Einstein equations.  As in the classical
theory, what we need is a few simple solutions which can be
interpreted and a degree of control on the structure of the {\it
space} of solutions.

We will conclude with a few remarks on the generalized Wick
transformation. First, note that $W$ does {\it not} arise from a
canonical transformation on the real phase space: the generating
function $T$ is imaginary, rather than real. (Thus, in quantum theory,
$\hat{W}$ is not expected to preserve norms.) As far as I can tell,
even on the complex phase space, the canonical transformation
generated by $T$ does not have a simple geometric property which can
readily enable one to interpret it directly as a ``Wick rotation''
from the Euclidean phase space to the Lorentzian. In particular, given
a specific solution to the {\it classical} Euclidean constraint, $W$
does not provide us with a solution to the Lorentzian constraint. It
is a well-defined mapping on the space of {\it functions} on the phase
space, rather than on the (real) phase space itself. Furthermore, in
general, it maps real functions to complex-valued functions; the
Hamiltonian constraint is more of an exception than a rule where a
real function is mapped to another real function. (For example, while
$f(x)\mapsto f(ix)$ will in general map real functions to complex,
$\cos x$ is again mapped to a real function, $\cosh x$.)  Next, the
action of $W$ does preserve the vector and the Gauss constraints
(modulo {\it overall} constants) so that if, as in the construction
sketched above, $\Psi$ were to satisfy all Euclidean constraints,
$\hat{W}\circ \Psi$ would satisfy {\it all} Lorentzian
constraints. Finally, one can ask whether, on the phase space, $W$
maps the Euclidean action to the Lorentzian. This is indeed the case
if one simultaneously transforms lapses and shifts via $(\ut{N},N^a)
\mapsto (-\ut{N}, N^a)$, where $\ut{N}$ is the lapse field with
density weight $-1$. These transformation properties of lapses and
shifts are, however, different from those one encounters in quantum
cosmology.

\vglue 0.6cm
\leftline{\twelvebf 6. Discussion}
\vglue 0.4cm

The new, background independent functional calculus on the quantum
configuration space $\agb$ has enabled us to develop the quantization
program systematically. The level of mathematical precision is such
that all underlying assumptions are explicit and we can be confident
that there are no hidden infinities. I would like to emphasize that,
for the problem under consideration, this degree of precision is not a
luxury; it is essential. Let me elaborate on this point since it is
often overlooked. In theoretical physics in general, and in quantum
field theory in particular, we often do formal manipulations, subtract
infinities and extract physical answers. We do not worry about
defining measures rigorously and are content even if the perturbation
series we arrive at diverges uncontrollably so long as individual
terms in the series are finite. In quantum gravity, however, I believe
that we can not afford to be so cavalier. For, the central problem is
somewhat different now. In the case of other three interactions, we
have piles of experimental data and the central task is that of
organizing it in a coherent fashion and making further predictions
that can be tested by experiments. In quantum gravity, we are not
blessed with this richness. At the present stage, the key problem is
that of consistency: Can the principles of general relativity and
quantum theory be unified in a mathematically consistent fashion? To
be confident of the answer, we are forced to elevate our mathematical
standards. Indeed, perturbative treatments have taught us that it
would be extremely unwise to be satisfied with formal manipulations.

The techniques underlying our background-independent functional
calculus may seem somewhat unfamiliar to most physicists.  The overall
situation is rather analogous to the introduction, in the sixties, of
global techniques in general relativity. Until then, most relativists
were content with local calculations and worked exclusively with
coordinates. When the various notions of causality and even the
definitions of a singularity were first introduced, they seemed exotic
to the practitioners in the field as they were outside of what was
then the mainstream.  The new techniques did not invalidate the use of
coordinate methods for local problems.  However, they turned out to be
indispensable to the analysis of global issues associated, e.g., with
horizons and singularities which were then emerging as the frontier
problems.  These were the types of problems that the older methods
could not handle. Indeed, using local methods, one could not even {\it
define} the notion of the event horizon of a black-hole much less
prove theorems about their properties. I believe that the situation
with the new functional calculus is quite analogous. These techniques
are essential to handle the question of mathematical consistency, the
frontier issue at the present stage of our understanding of quantum
gravity. Traditional tools, powerful as they are in the analysis of
other interactions, appear to be insufficient to face this issue
non-perturbatively, without any recourse to a background geometry.
With the new calculus, we have tools to meet this challenge.

As we saw, these techniques have already been applied to two problems
with considerable success: probing quantum geometry and solving the 
quantum constraints of the general relativity. 

The picture of the geometry at the Planck scale that emerged turned
out to be very different from the continuum image. In particular,
operators representing areas of 2-surfaces and volumes of
3-dimensional regions have discrete spectra. This departure from the
continuum --and particularly the discreteness of the spectrum of the
area operators--may well hold the key to some of the current puzzles
such as the relation between statistical entropy and the area of black
hole horizons [43].  The fundamental, Planck scale excitations of the
gravitational field are 1-dimen\-sion\-al and the corresponding
geometry is distributional.  As was pointed out in Section 3, in this
respect, there is a close similarity with polymer physics.  Although
polymers and the basic phonon excitations in them are 1-dimensional,
in suitably complex configurations, they exhibit 3-dimensional
properties. The same is true of geometry [3,44]: continuum
3-dimensional geometries {\it do} arise, but as approximations. More
precisely, the Hilbert space $H_o$ admits states which can be
interpreted as ``semi-classical'' in the sense that, when
coarse-grained using a macroscopic length scale, they become
indistinguishable from smooth continuum geometries in three
dimensions. For example, given a flat 3-metric $g^o_{ab}$ on $R^3$,
and a macroscopic length scale $L$, we can ask for states $\Psi$ in
$H_o$ which have the property that for all regions with $g^o$-volume
of the order of $L^3$ or bigger and for all surfaces $S$ which are
slowly varying on the scale $L$, the eigenvalues of the volume and
area operator approximate the values one would obtain {\it
classically} using $g^o_{ab}$, up to corrections of the order
$O(\ell_P/L)$. Such states exist and can be constructed using
techniques from ``statistical geometry'' that underlie random lattices
[44]. These states have been generically called ``weaves'' because
they tell us how to weave a classical geometry from ``quantum
threads'' --the elementary excitations of geometry.

There is an indirect test of these ideas. Although one expects the
laboratory physics to be insensitive to the detailed predictions of
quantum gravity, it may well be that to do this physics in a
coherent fashion, one has to supplement the standard description with
some {\it qualitative} ideas from quantum gravity%
\footnote{${}^6$}{\ninerm \baselineskip=5pt\noindent There are 
numerous examples of such situations in other branches of physics. For
example, in astrophysics, one can generally work with Newtonian
gravity.  However, to discuss the density distributions of stars near
the centers of galaxies such as M87 which have large black holes in
their centers, one simply changes the boundary conditions on Newtonian
equations, allowing stars to disappear once they cross the event
horizons. Thus, although the use full general relativity is
unnecessary, some ``qualitatively new'' features of this more accurate
theory have to be incorporated.}. %
For instance, in Minkowskian quantum field theories, the main
difficulty comes from the ultra-violet divergences which arise because
we integrate over arbitrarily large momenta, i.e., arbitrarily short
distances. A more accurate representation of the underlying geometry
is given by a weave state. So, rather than doing quantum field theory
on a continuum, we should really do it on a weave state. Such a theory
would be free of ultra-violet divergences. The key question is whether
it reproduces the results of the standard perturbation theory at ``low
energy.''  Heuristically, one would expect this to be the case: Since
the standard theories are renormalizable, their predictions for
phenomena at the $10^{-17}$cm scale should be insensitive to the
details of the micro-structure of weave states at the Planck
scale. (Recall that the weave geometry differs from the continuum only
up to terms $O(\ell_P/L)\approx (10^{-33}/10^{-17}) \approx 10^{-16}$.)
If one could show this result in detail  --i.e., establish that the
predictions of a {\it finite} quantum field theory on a weave state
agree with those of the standard perturbation theory for laboratory
energies-- one could take the rich phenomenological data from particle
physics as an indirect evidence for the discrete structure of the
Planck scale geometry.

The second main application of the framework is to quantum Einstein
equations.  The diffeomorphism constraints could be regulated on the
kinematical Hilbert space $H_o$ in a manner that is anomaly-free and
the space of solutions could be given a Hilbert space structure using
the group averaging technique.  There is also progress on the
Hamiltonian constraint and the idea of using a generalized Wick
transformation is tantalizing. However, more work is needed to make
these results definitive. It is quite possible that to obtain a
satisfactory regularization, one would have to bring in the notion of
``framed networks'' and replace the gauge group $SU(2)$ by its
``quantum version''. However, if these last problems could be
resolved, one would conclude that quantum general relativity does
exist non-perturbatively. It is already clear that its structure would
be very different from that envisaged in perturbative treatments which
are all rooted in a continuum picture. My own view is that there may
well exist several inequivalent ways of regulating the Hamiltonian
constraint, leading to inequivalent theories. Indeed, this is likely
to be physically the most important source of non-uniqueness of the
quantum theory. If the theory does admit such inequivalent sectors, it
would be all the more important to develop approximation schemes to
extract their physical content. They will be motivated by the
structure of the exact theory and therefore likely to be quite
different from the ones used in the standard perturbative
treatments. Work is in progress along several lines in this area.

An attractive possibility is to combine the strengths of string theory
and the non-perturbative approach discussed here. String theory
provides a ``tight'' strategy to couple matter for which no analogous
principle is known within general relativity. On the other hand, at
least in practice, string theory is essentially perturbative and needs
a background continuum geometry. Quantum general relativity needs no
background fields and provides a specific picture of quantum geometry.
A natural strategy then would be to investigate strings on these
quantum geometries. Indeed, since the excitations of geometry are
along 1-dimensional graphs, it it natural to incorporate matter
through strings winding around loops in these graphs. This might
provide for us a viable perturbation theory. Results of Klebanov and
Susskind [45] suggest a concrete direction for this work.

To conclude, let me address an obvious question about the new
functional calculus: Since it refers to theories of connections, can
it not be used for a non-perturbative treatment of Yang-Mills
theories? Unfortunately, in the general case, there is an obstruction:
While our emphasis has been on diffeomorphism invariance and absence
of background fields, Yang-Mills theories depend on the background
Minkowskian metric rather heavily. Thus, to make the framework
directly amenable to {\it general} Yang-Mills theories, we would have
to develop it further in a substantial way by adding new techniques
which are tailored to the presence of a flat background metric. There
is, however, an exception: two space-time dimensions where one only
needs an area element, rather than a metric, to specify the Yang-Mills
action. Thus, the theories are now invariant under area preserving
diffeomorphisms. In this case, our techniques {\it are} directly
applicable and have led to a number of {\it new} results [46]. These
include: explicit expressions of the Schwinger functions for Wilson
loops and a direct proof of the equivalence of the Euclidean path
integral formulation and the Hamiltonian quantum theory. Furthermore,
unlike in other approaches, the invariance of the quantum theory under
area preserving diffeomorphisms is manifest in this treatment.

\vglue 0.6cm
\leftline{\twelvebf Acknowledgements}
\vglue 0.4cm

Most of the recent results reported here were obtained in
collaboration with C. Isham, D. Marolf, J. Mour\~ao, T. Thiemann and,
especially, J. Lewandowski.  I am most grateful to them for constant
intellectual stimulation over the past two years and for their
patience with my slow pace in writing up the results.  I am grateful
also to John Baez, Rodolfo Gambini, Karel Kucha\v{r}, Jorge Pullin,
Carlo Rovelli, and Lee Smolin for their numerous suggestions, comments
and criticisms which were often vital.  This work was supported in
part by the NSF grant PHY 93-96246 and by the Eberly Research Funds of
the Penn State University.
\goodbreak

\vglue 0.6cm
\leftline{\twelvebf References}
\vglue 0.4cm

\itemitem{1.} P. Faria de Veiga, Ecole Polytechnique thesis (1990);
C. de Calan, P. Faria de Veiga, J. Magnen and R. S\'en\'eor, {\tit
Phy. Rev. Lett.} {\tbf 66} (1991) 3233; A. S. Wightman, in: {\twelveit
Mathematical Physics Towards XXIst Century}, eds R. N. Sen and
A. Gersten (Ben Gurion University Press, 1994).
\itemitem{2.} D. Amati, M. Ciafolini and G. Veneziano, {\tit Nucl. Phys.}
{\tbf B347} (1990) 550.
\itemitem{3.} A. Ashtekar, C. Rovelli and L. Smolin, {\tit Phys. Rev. 
Lett.} {\tbf 69} (1992) 237; 
\itemitem{4.} A. Agishtein and A. Migdal, {\twelveit Mod. Phys. Lett.}  
{\tbf 7} (1992) 85.
\itemitem{5.} J. Iwasaki and C. Rovelli, {\tit Int. J. Mod. Phys.}
{\tbf D1} (1993) 533; {\tit Class. \& Quantum Grav.}{\tbf 11} (1994)
1653.
\itemitem{6.} T. Jacobson and L. Smolin, {\tit Nucl. Phys.} {\tbf B299}
(1988) 295.
\itemitem{7.} C. Rovelli and L. Smolin, {\tit Phys.  Rev. Lett.} 
{\tbf 72} (1994) 446.
\itemitem{8.}A. Ashtekar, {\tit Non-Perturbative Canonical Gravity}
(World Scientific, Singapore, 1991); in {\tit Gravitation and
 Quantization} eds  B. Julia and j. Zinn-Justin (Elsevier, Amsterdam
 1995).
\itemitem{9.} R. Rovelli, {\tit Class. \& Quantum Grav.} {\tbf 8} (1991)
1613.
\itemitem{10.} L. Smolin, in {\tit Quantum Gravity and Cosmology}
eds J. P. Mercader, H. Sol\`a and E. Verdaguer (World Scientific,
Singapore, 1992).
\itemitem{11.} R. Gambini and J. Pullin, {\tit Loops, Knots, Gauge 
Theories and Quantum Gravity} (Cambridge University Press, Cambridge, 1996).
\itemitem{12.} A. Ashtekar, {\tit Phys. Rev. Lett.} {\tbf 57} (1986) 2244;
{\tit Phys. Rev.}{\tbf D36} (1987) 1587.
\itemitem{13.} A. Ashtekar, in {\tit Mathematics and General Relativity}
(AMS, Providence, \- 1987), J.F. Barbero G. {\tit Phys. Rev} {\tbf D51}
(1995) 5507.
\itemitem{14.} A. Ashtekar and C. J. Isham, {\tit Class. \& Quantum Grav.} 
{\tbf 9} (1992) 1433.
\itemitem{15.} J. C. Baez and S. Sawin, {\tit Functional integration of 
spaces of connections}, q-alg/9507023.
\itemitem{16.} A. Rendall, {\tit Class. \& Quantum Grav.} {\tbf 10} (1993)
605.   
\itemitem{17.} A. Ashtekar and J. Lewandowski, in {\tit Knots and Quantum
Gravity}, ed J. Baez (Oxford University Press, Oxford, 1994).
\itemitem{18.} J. Baez, {\tit Lett. Math. Phys.}{\tbf 31} (1994) 213;  
in {\tit The Proceedings of the Conference on Quantum Topology},
ed D. N. Yetter (World Scientific, Singapore, in press).
\itemitem{19.} D. Marolf and J. Mor\~ao, {\tit Commun. Math. Phys.}
{\tbf 170} (1995) 583.
\itemitem{20.} A. Ashtekar and J. Lewandowski, {\tit J. Math. Phys.}
{\tbf 36} (1995) 2170.
\itemitem{21.} A. Ashtekar and J. Lewandowski, {\tit J. Geo. \& Phys.}
{\tbf 17} (1995) 191.
\itemitem{22.} A. Ashtekar, J. Lewandowski, D. Marolf, J. Mour\~ao and 
T. Thiemann, {\tit Coherent state transform on the space of connections},
{\tit J. Funct. Analysis} (in press).
\itemitem{23.} A. Ashtekar, J. Lewandowski, D. Marolf, J. Mour\~ao and 
T. Thiemann, {\tit J. Math. Phys.}{\tbf 36} (1995) 6456.
\itemitem{24.} J.  C. Baez, {\tit Spin networks in gauge theory}, 
{\tit Adv. Math.} (in press); {\tit Spin networks in non-perturbative
quantum gravity}, gr-qc/9504036.
\itemitem{25.} L. Smolin (private communication).
\itemitem{26.} C. Rovelli and L. Smolin, {\tit Spin networks and quantum
gravity}, preprint CGPG-95/4-1. 
\itemitem{27.} R. Penrose, in {\tit Quantum Theory and Beyond}, ed 
T. Bastin, Cambridge University Press, Cambridge 1971).
\itemitem{28.} A. Ashtekar and J. Lewandowski, {\tit Quantum
geometry,} (preprint).
\itemitem{29.} C. Rovelli and L. Smolin, {\tit Nucl. Phys.}{\tbf B442},
593 (1995).
\itemitem{30.} R. Loll, {\tit The volume operator in discretized
gravity,} (pre-print).
\itemitem{31.} C. Rovelli, {\tit Nucl. Phys.}{\tbf B405} (1993) 797;
L. Smolin, {\tit Phys. Rev.}{\tbf D49} (1994) 4028.
\itemitem{32.} A. Higuchi, {\tit Class. \& Quantum Grav.}{\tbf 8} (1991)
1983; 2023; N. P. Landsman {\tit J. Geo. \& Phys.}{\tbf 15} (1995) 285.
\itemitem{33.} A. Ashtekar and R. S. Tate, {\tit J. Math. Phys.} 
{\tbf 34} (1994) 6434.
\itemitem{34.} B. Br\"ugmann and J. Pullin, {\tit Nucl. Phys.} {\tbf B363}
(1991) 221; {\tbf B390} (1993) 399. 
\itemitem{35.} B. Br\"ugmann, R. Gambini and J. Pullin, {\tit Phys. Rev.
Lett.} {\tbf 68} (1992) 431.
\itemitem{36.} H. Nicolai and H. J. Matschull, {\tit J. Geo. and Phys.}
{\tbf 11} (1993) 15; H. J. Matschull, pre-print gr/qc 9305025;
\itemitem{37.} H. A. Morales-T\'ecotl and C. Rovelli, {\tit Phys. Rev. 
Lett.}{\tbf 72} (1994) 3642. 
\itemitem{38.} C. Rovelli and L. Smolin, (private communication).
\itemitem{39.} A. Ashtekar and J. Lewandowski (in preparation).
\itemitem{40.} K.  Kucha\v{r} (private communication).
\itemitem{41.} T. Thiemann, {\tit Reality conditions inducing transforms 
for quantum gauge fields and quantum gravity,} (pre-print).
\itemitem{42.} A. Ashtekar, {\tit A generalized Wick transform for gravity,}
(pre-print).
\itemitem{43.} J. D. Bekenstein and V. F. Mukhanov, {\tit Spectroscopy 
of quantum black holes,} (preprint).
\itemitem{44.} A. Ashtekar and L. Bombelli (in preparation).
\itemitem{45.} I. Klebanov and L. Susskind, {\tit Nucl. Phys.} 
{\tbf B309} (1988) 175. 
\itemitem{46.} A. Ashtekar, J. Lewandowski, D. Marolf, J. Mora\~o and 
T. Thiemann, in {\tit Geometry of Constrained Dynamical Systems,}
ed. J.  Charap (Cambridge University Press, Cambridge, 1994); {\tit
Quantum Yang-Mills theory in two dimensions: A complete solution,}
(pre-print).

\end\bye